\documentclass[
 reprint,
superscriptaddress,
 amsmath,amssymb,
 aps,
prl,
]{revtex4-2}

\usepackage{graphicx}
\usepackage{dcolumn}
\usepackage{bm}
\usepackage{mathrsfs}
\usepackage{csquotes}
\usepackage{hyperref}
\hypersetup{
    colorlinks=true,
    citecolor=blue,      
    linkcolor=blue,
}

\usepackage{calligra}
\DeclareMathAlphabet{\mathcalligra}{T1}{calligra}{m}{n}
\DeclareFontShape{T1}{calligra}{m}{n}{<->s*[2.2]callig15}{}

\makeatletter
\let\@orig@make@capt@title\@make@capt@title
\def\@make@capt@title#1#2{\@orig@make@capt@title{{\bf #1}}{#2}}
\makeatother

\begin{document}

\title{Viscous tweezers: controlling particle orientation with viscosity}

\author{Tali Khain}
\affiliation{James Franck Institute and Department of Physics, The University of Chicago, Chicago, IL 60637, USA}
\author{Michel Fruchart}
\affiliation{James Franck Institute and Department of Physics, The University of Chicago, Chicago, IL 60637, USA}
\affiliation{Gulliver, UMR CNRS 7083, ESPCI Paris PSL, 75005 Paris, France}
\author{Vincenzo Vitelli}
\affiliation{James Franck Institute and Department of Physics, The University of Chicago, Chicago, IL 60637, USA}
\affiliation{Kadanoff Center for Theoretical Physics, The University of Chicago, Chicago, IL 60637, USA}

\begin{abstract}
Control of particle motion is generally achieved by applying an external field that acts directly on each particle.
Here, we propose a global way to manipulate the motion of a particle by dynamically changing the properties of the fluid in which it is immersed. 
We exemplify this principle by considering a small particle sinking in an anisotropic fluid whose viscosity depends on the shear axis.
In the Stokes regime, the motion of an immersed object is fully determined by the viscosity of the fluid through the mobility matrix, which we explicitly compute for a pushpin-shaped particle.
Rather than falling upright under the force of gravity, as in an isotropic fluid, the pushpin tilts to the side, sedimenting at an angle determined by the viscosity anisotropy axis.
By changing this axis, we demonstrate control over the pushpin orientation as it sinks, even in the presence of noise, using a closed feedback loop.
This strategy to control particle motion, that we dub viscous tweezers, could be experimentally realized in systems ranging from polyatomic fluids under external fields to chiral active fluids of spinning particles by suitably changing their direction of global alignment or anisotropy.
\end{abstract}

\maketitle

The control of small particles in a fluid is crucial in applications ranging from crystal growth~\cite{Boles2016} to drug delivery~\cite{Nelson2010,Walker2022}.
It arises in diverse contexts such as sedimentation~\cite{Ramaswamy2001,Guazzelli2009}, crystalization~\cite{Boles2016}, swimming~\cite{Lauga2009}, or self-propulsion~\cite{Bär2020,
Gompper2020,Bechinger2016,Moran2017}. 

To control the state of a single particle, it is common to apply external fields that act on the particle by enacting a force or a torque~\cite{Walker2022}, either directly or through externally imposed gradients~\cite{Anderson1989,Moran2017}.
For instance, optical tweezers work because dielectric particles are pushed upwards gradients of light intensity~\cite{svoboda1994biological,moffitt2008recent}. 
Similar effects arise in gradients of other fields such as sound intensity~\cite{ozcelik2018acoustic} or chemical concentration~\cite{Anderson1989}. 
However, these effects only work on particles that can respond to the imposed field, in the same way as dielectrics respond to an electric field by developing a polarisation.

A way to act universally on broader classes of particles consists in directly harnessing the properties of the fluid in which the particles are immersed.
For instance, it has been reported that spatial gradients in viscosity can be used to guide particles in space through an effect known as viscotaxis~\cite{oppenheimer2016motion, liebchen2018viscotaxis, shoele2018effects, datt2019active,laumann2019focusing, eastham2020axisymmetric,dandekar2020swimming, shaik2021hydrodynamics, lopez2021dynamics, anand2024sedimentation, gong2023active, gong2024active}.
In parallel, in nematic liquid crystals, the elasticity of the order parameter -- the tendency of the nematic building blocks to align with each other combined with their tendency to be aligned (or \enquote{anchored}) in a certain way at particle boundaries -- can be harnessed to control particle motion through the orientation of the nematic director~\cite{smalyukh2018liquid,senyuk2013rotational, chi2020surface, chandler2023nematic,lapointe2009shape, martinez2011large, silvestre2014towards, yoshida2015three, li2017directed, tkalec2011reconfigurable, luo2018tunable}.
Similar results have been developed for other classes of liquid crystals~\cite{smalyukh2018liquid}.

In this Letter, we show that manipulating the tensorial structure of the viscosity of an anisotropic fluid implements a way to indirectly control the motion or orientation of an immersed object subjected to a constant force.
This method is independent of the nature of the particle and does not impose a predetermined flow in the fluid.
The basic requirement, a fluid with a tunable axis of anisotropy, is present in systems such as polyatomic fluids under electric or magnetic fields \cite{Beenakker1970}, electron fluids \cite{Varnavides2020,Gusev2020,Cook2021,Moll2016}, chiral active fluids of spinning particles~\cite{soni2019odd, markovich2021odd, han2021fluctuating, khain2022stokes, fruchart2023odd}, or so-called viscosity metamaterials, which are complex fluids whose viscosity can be controlled by applying acoustic perturbations \cite{sehgal2019using,sehgal2022viscosity,Gibaud2020}.
In contrast with viscotaxis, our method does not require gradients of viscosity.
In addition, it does not rely on the existence of nematic elasticity in the fluid, nor anchoring boundary conditions at the surface of the particle, as it acts only through viscous effects.
Hence, we expect it to work in any anisotropic fluid, including those like polyatomic fluids~\cite{Beenakker1970} and electron fluids in anisotropic solids~\cite{Varnavides2020,Gusev2020,Cook2021,Moll2016}, where constituents can be aligned by an external field but (unlike nematic liquid crystals) do not have a strong tendency to align with each other.

\begin{figure*}
    \centering
    \includegraphics[width=\textwidth]{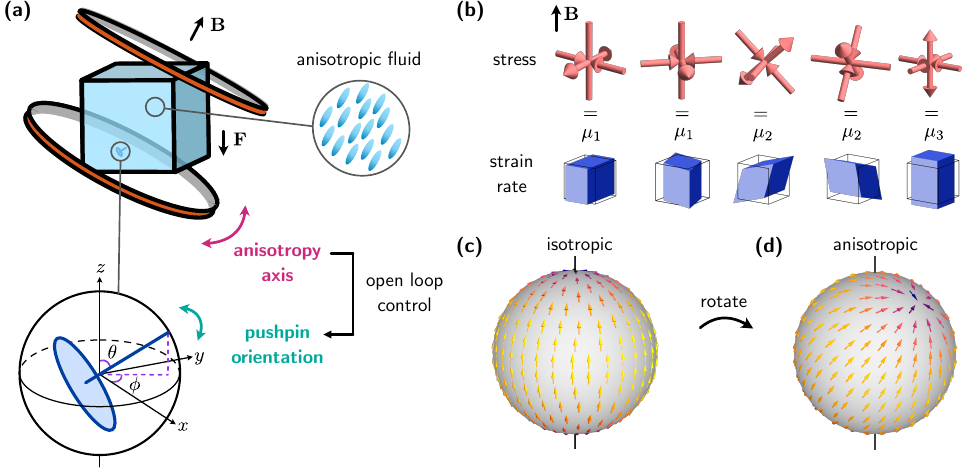}
    \caption{\label{fig:setup} 
    \textbf{Controlling the orientation of a pushpin sinking in an anisotropic fluid by changing the viscosity axis.} 
    (a) A pushpin sinking under the force of gravity in a tank filled with fluid that is under a magnetic field (top left). 
    The fluid is anisotropic, as it consists of elongated molecules that align along the direction of $\bm{B}$ (top right). 
    The pushpin is a cylindrically symmetric object, and so its orientation can be described by two angles, $\theta$ and $\phi$ (bottom left). 
    A given orientation of the pushpin corresponds to a point on the surface of a sphere. 
    By changing the direction of the magnetic field $\bm{B}$ (and thus the anisotropy axis of the fluid), we change the orientation of the pushpin as it sinks (bottom right). 
    (b) A fluid that microscopically obeys cylindrical symmetry about an axis, which we call $\bm{B}$, can have three shear viscosities, $\mu_1, \mu_2,$ and $\mu_3$. The stress and strain rate deformations associated with each of these viscosities are pictorially shown for $\hat{\bm{B}} = \hat{\bm{z}}$. 
    If the anisotropy axis is rotated, as in (a), so does the viscosity.
    (c) The orientation dynamics of a pushpin under a vertical force in an isotropic fluid, visualized on the surface of a sphere. The orientation flows towards a fixed point at the north pole, as the pushpin rights itself up to sink vertically. 
    (d) In an anisotropic fluid, like the one in (a), one could imagine that the flow would be a rotated version of the one in (c). 
    In this case, the fixed point could shift off the north pole, causing the pushpin to fall at an angle.}
\end{figure*}

\textit{Stokes flow and mobility.}---Let us consider a small rigid particle immersed in a viscous incompressible fluid. 
In this low Reynolds number regime, the fluid flow is well-described by the Stokes equation,
\begin{align}
    \partial_t v_i = \partial_j \sigma_{ij} + f_i
    \quad
    \text{with}
    \quad
    \sigma_{ij} = -\delta_{ij} P + \eta_{ijk\ell} \partial_\ell v_k
\end{align}
along with the incompressibility condition $\partial_i v_i = 0$. 
Here, $v_i$ is the fluid velocity, $P$ the pressure, $\sigma_{ij}$  the stress tensor, $f_i$ an external force, and $\eta_{ijk\ell}$ the viscosity tensor. 
The overdamped motion of a particle in a fluid is described by the linear equation
\begin{align}
    \begin{bmatrix}
    \bm{V}\\
    \bm{\Omega}
    \end{bmatrix}
    = 
    \mathbb{M}(\eta)
    \begin{bmatrix}
    \bm{F}\\
    \bm{\tau}
    \end{bmatrix}
    \label{eq:mobility},
\end{align}
where the $6 \times 6$ mobility matrix $\mathbb{M}$ relates the force $\bm{F}$ and torque $\bm{\tau}$ applied to the particle with its velocity $\bm{V}$ and angular velocity $\bm{\Omega}$ \cite{KimKarrila, khain2024trading}. 
The form of $\mathbb{M}$ depends on both the geometry of the object and the viscosity tensor $\eta$ of the fluid.
The position and orientation of the particle can then be obtained by integrating the velocity and angular velocity.

We focus on the orientation of a sedimenting particle that sinks under the force due to gravity $\bm{F} = F \hat{\bm{z}}$. Note that here we apply no torque ($\bm{\tau} = \bm{0}$), which is the most common way of changing the orientation of the particle. Equation~\eqref{eq:mobility} then reduces to
\begin{align}
\bm{\Omega} = T(\eta) \bm{F}
\label{eq:mobility_T}
\end{align}
in which $T$ is a sub-block of $\mathbb{M}$, see Appendix~\hyperref[app:Stokeslet_approx]{E}. As the force and the object are given, our only handle on the orientation dynamics is the viscosity tensor $\eta$ in Eq.~\eqref{eq:mobility_T}.

\begin{figure*}
    \centering
    \includegraphics[width=\textwidth]{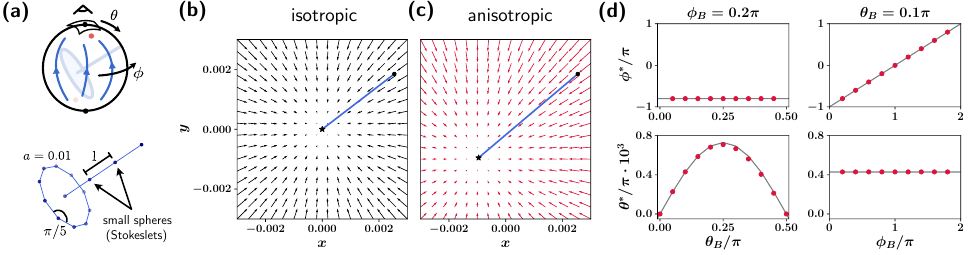}
    \caption{\label{fig:numerics} 
    \textbf{Orientation dynamics in an anisotropic fluid.} (a) The phase portrait of the orientation of the pushpin can be visualized on the surface of a sphere (top panel). In an isotropic fluid, the fixed points are at the north and south poles (black). In an anisotropic fluid, the fixed points can shift (red). The pushpin itself is constructed out of small spheres, or Stokeslets (bottom panel). The spheres cannot move with respect to one another, forming a rigid object.  (b) The flow near the north pole, as visualized from above, in an isotropic fluid. The blue line shows a trajectory of the pushpin orientation as it approaches the fixed point at the origin. (c) In an anisotropic fluid, the fixed point shifts off of the origin. Here, $\epsilon = 0.1, \theta_B = 0.1\pi, \phi_B = 0.25\pi$. (d) The position of the fixed point ($\phi^*,\theta^*$) as a function of the magnetic field (viscosity) axis ($\theta_B, \phi_B$), for $\epsilon = 0.1$.}
\end{figure*}

\textit{Viscosity of an anisotropic fluid.}---In familiar fluids such as water, this viscosity tensor reduces to one scalar coefficient, the shear viscosity $\mu$.
When the fluid is anisotropic (for example, a fluid consisting of elongated molecules that are aligned to an externally applied magnetic field, $\bm{B}$, as in Fig.~\ref{fig:setup}a), the shear viscosity of the fluid may not be the same in all directions, but depends on the shear axis. 
Assuming that the viscosity tensor is invariant under rotations about the anisotropy (alignment) axis, the most general equation of motion can contain three shear viscosities (see Appendix~\hyperref[app:Stokes_general]{A}).
The shear stress and strain rate deformations corresponding to these viscosities are visualized in Fig.~\ref{fig:setup}b, for an anisotropy axis chosen along the $z$ direction.
Here, we consider Newtonian viscosities, which are independent of the local velocity field and are only a function of the externally applied field, $\bm{B}$.

In a generic fluid, the magnitude of the anisotropy could depend on both $\bm{B}$ and on the microscopic details of the system.
Here, we separate out the orientation and magnitude: $\hat{\bm{B}}$ controls the direction of the anisotropy axis and $\epsilon$ sets the strength of the anisotropy.
In this case, the Stokes equation is
\begin{align}
-\frac{1}{\mu} \nabla P + \Delta \bm{v} + \epsilon \mathcal{D}(\hat{\bm{B}})\bm{v} = 0,
\label{eq:Stokes_anisotropic}
\end{align}
where $\mathcal{D}$ is a matrix of second derivatives. 
As an example, consider a weakly anisotropic fluid with shear viscosities $\mu_1 = \mu$, $\mu_2 = \mu(1 + \epsilon)$, and $\mu_3 = \mu(1 + \frac{4}{3}\epsilon)$ when the anisotropy axis is along the $z$ direction (see Fig.~\ref{fig:setup}b and Appendix~\hyperref[app:Stokes_general]{A}). This particular form allows for analytical calculations when $\epsilon$ is small (Appendix~\hyperref[app:Greens_function]{B}-\hyperref[app:drag_sphere]{D}), but our general strategy applies to any anisotropic viscosity. The operator $\mathcal{D}$ then takes the form
\begin{align}
\mathcal{D}(\hat{\bm{B}} = \hat{\bm{z}}) = 
    \begin{bmatrix}
        \partial_z^2 & 0 & - \partial_x \partial_z / 3 \\
        0 & \partial_z^2 & -\partial_y \partial_z / 3 \\
        0 & 0 & \Delta + 2\partial_z^2 /3
    \end{bmatrix}
    \label{eq:Stokes_Bz}
\end{align}
The matrix $\mathcal{D}$ could be chosen to be symmetric by using incompressibility.

The Green function of the Stokes equation (Stokeslet) can be computed numerically for any value of $\epsilon$ using fast Fourier transforms. We compute it analytically in the perturbative regime to linear order in $\epsilon$
(Appendix~\hyperref[app:Greens_function]{B}).

\textit{Motion of a pushpin in an anisotropic fluid.}---To determine the form of $\mathbb{M}$, we now need to specify the shape of our particle.
In principle, this requires solving boundary value problems for this specific shape~\cite{KimKarrila}. We use a shortcut by which the mobility matrix for a given shape is obtained by constructing the object out of Stokeslets (see Appendix~\hyperref[app:Stokeslet_approx]{E} and Refs.~\cite{witten2020review,mowitz2017predicting}).
To validate this method, we first consider a sphere. In this case, we can analytically solve the boundary value problem of a fluid flowing past the sphere in the limit of weak anisotropy (small $\epsilon$), calculate the force and torque that the fluid exerts on the object, and compare with the results of the Stokeslet method (Appendix~\hyperref[app:Stokeslet_approx]{E}).
The main consequence is that a sphere settling under the force of gravity in an anisotropic fluid sinks slower than in an isotropic one. The familiar Stokes drag law is modified: the drag coefficient is increased in the $x$ and $y$ directions by a factor of $(1 + \epsilon/2)$ and in the $z$ direction by $(1 + \epsilon)$.

When the shape of the particle is not spherically symmetric, both its velocity and angular velocity can change as compared to the isotropic case. 
We consider the simplest shape which exhibits non-trivial orientation evolution: a cylindrically symmetric pushpin, shown in Fig.~\ref{fig:setup}a. The orientation of the pushpin is described by two angles: $\theta$, the angle the pushpin long axis makes with the lab $z$ axis, and $\phi$, the angle between the plane projection of the pushpin long axis and the lab $x$ axis. Equivalently, the pushpin orientation is given by the radial unit vector
\begin{align}
\hat{\bm{n}}(\theta,\phi) = (\sin{(\theta)} \cos{(\phi)}, \sin{(\theta)}\sin{(\phi)}, \cos{(\theta)}).
\end{align}
The mobility matrix $\mathbb{M} = \mathbb{M} (\epsilon, \hat{\bm{B}}, \hat{\bm{n}})$, which determines how the pushpin moves, depends on the orientation $\hat{\bm{n}}$ of the pushpin, on the anisotropy axis $\hat{\bm{B}} = (\cos{(\phi_B)}\sin{(\theta_B)}, \sin{(\phi_B)}\sin{(\theta_B)},\cos{(\theta_B}))$ of the fluid (Eq.~\ref{eq:Stokes_Bz} is written with $\hat{\bm{B}} = \hat{\bm{z}}$), and on the strength of the anisotropy $\epsilon$. By constructing the pushpin out of rigidly-connected Stokeslets (see bottom panel of Fig.~\ref{fig:numerics}a), we can compute the mobility matrix for any anisotropy direction and pushpin orientation (see Appendix~\hyperref[app:Stokeslet_approx]{E} for more details).
Examples of mobility matrices for a tilted pushpin in an isotropic and anisotropic fluid can be visualized schematically as 
\begin{equation}
\hspace{-0.9cm}
    \raisebox{-0.5\totalheight}{\includegraphics[width=\columnwidth]{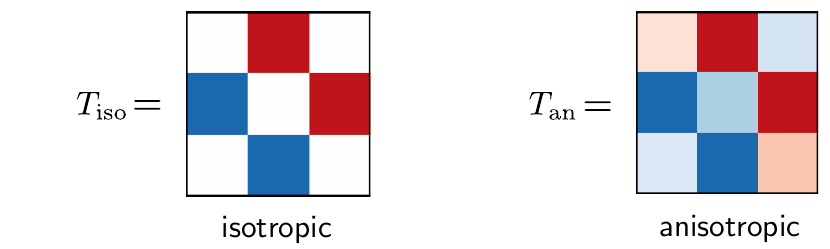}} 
    \label{eq:M_orient}
\end{equation}
in which red/blue represent positive/negative entries whose magnitude is represented by lightness (see Appendix~\hyperref[app:Stokeslet_approx]{E}).

\begin{figure}
    \centering
    \includegraphics[width=\columnwidth]{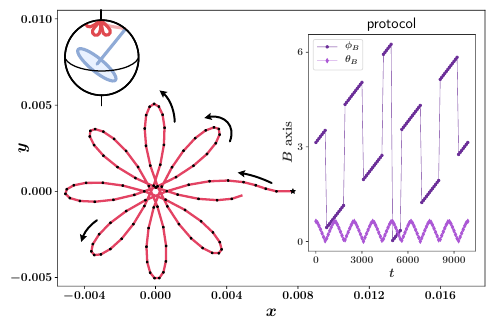}
    \caption{\label{fig:control} 
    \textbf{Demonstration of control of pushpin orientation with viscosity.} The pushpin can be made to follow any orientation trajectory by changing the magnetic field axis. The pictured rose plot corresponds to the trajectory of the pushpin orientation near the north pole ($\theta = 0$), which is generated by the protocol in the inset.}
\end{figure}

\begin{figure*}
    \centering
    \includegraphics[width=\textwidth]{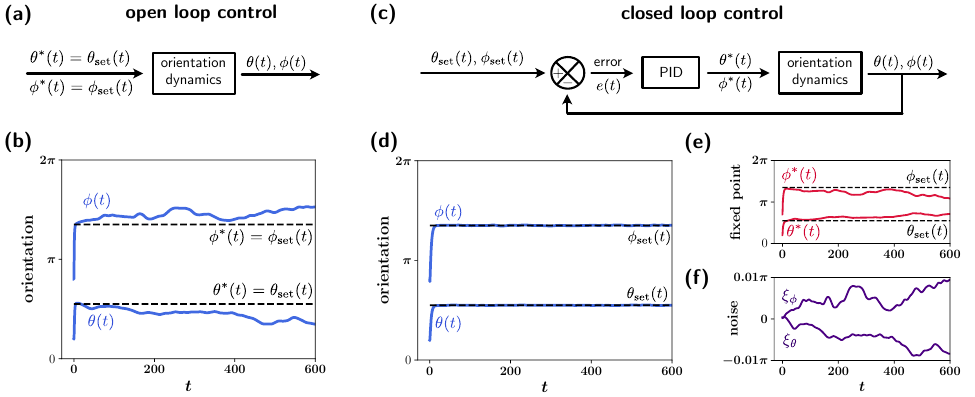}    \caption{\label{fig:control_loops} 
    \textbf{Open loop vs. closed loop control in toy model of pushpin orientation.} (a-b) Open loop control diagram for the orientation toy model. The fixed point ($\theta^*, \phi^*$) is set to the target orientation ($\theta_{\text{set}}, \phi_{\text{set}}$) (dashed black curves). In the presence of noise $\xi$ (f), the orientation (blue) evolves to hover near the target with a visible error. (c-e). Closed loop diagram with a PID controller for the orientation toy model. In this case, the orientation successfully adheres to the set target, despite noise (f). Here, the fixed point $(\theta^*, \phi^*)$ is no longer set, but is evolved by the PID controller in response to the noise (e). For the PID controller parameters, we use $K_p = 0.4 \pi, K_i = 0.2\pi, K_d = 0$.}
\end{figure*}

\textit{Orientation dynamics of a sedimenting pushpin.}---We now investigate the dynamics of a pushpin sinking under the force of gravity. 
Applying a constant force in the $-z$ direction determines the angular velocity $\bm{\Omega}$ of the pushpin, as in Eq.~\ref{eq:mobility_T}. 
Then, the equation of motion for the orientation of the pushpin is given by
\begin{equation}
    \partial_t \bm{\hat{n}} = \bm{N}(\bm{\hat{n}}) \equiv \bm{\Omega} \times \bm{\hat{n}}
    \label{eq:N}
\end{equation}
in which $\bm{\Omega}$ is given by Eq.~\eqref{eq:mobility_T}.
Since $\bm{N}\cdot \hat{\bm{n}} = 0$, the vector field $\bm{N}$ describing the orientation dynamics of the pushpin is tangent to the sphere (there is no radial component), as shown in Fig.~\ref{fig:setup}c.
The arrows show the instantaneous motion of the tip of a pushpin embedded in the center of the sphere.
In spherical coordinates, Eq.~\ref{eq:N} reads
\begin{subequations}
\label{eq:eom_spherical}
\begin{align}
\dot{\theta} = N_{\theta} \label{eq:eom_theta} \\
\sin{(\theta)}\dot{\phi} = N_{\phi},
\label{eq:eom_phi}
\end{align}
\end{subequations}
which we numerically solve with an explicit Runge-Kutta method of order 5(4) as implemented in SciPy \texttt{solve\_ivp} \cite{scipy}.

We now ask what is the eventual orientation of the pushpin. Fixed points of the orientation dynamics satisfy
\begin{equation}
\bm{N} (\theta^*, \phi^*) = 0.
\end{equation}
In the isotropic case ($\epsilon=0$), we find that after a transient, the pushpin orients itself to fall upright, with $\theta=0$ (Fig.~\ref{fig:setup}c). 
We expect that the anisotropy in the direction $\bm{B}$ will tilt the pushpin at an angle depending on the anisotropy direction and strength (Fig.~\ref{fig:setup}d). 
Such a setup would allow us to control the orientation of the pushpin by acting on the fluid (Fig.~\ref{fig:setup}a). We confirm that this is indeed the case using numerical simulations of the orientation dynamics.

The results of our numerical simulations are presented in Fig.~\ref{fig:numerics}, in which we zoom in on the region of the sphere around the north pole, which corresponds to the stable fixed point in an isotropic fluid (Fig.~\ref{fig:numerics}a-b). 
In the anisotropic case ($\epsilon \neq 0$), the steady state orientation of the pushpin can change: in Fig.~\ref{fig:numerics}c, the fixed point moves off of the north pole.

We numerically compute the dependence of the fixed point position ($\theta^*, \phi^*$) on the orientation of the anisotropy axis ($\theta_B, \phi_B$) in Fig.~\ref{fig:numerics}d. 
As long as the anisotropy axis $\hat{\bm{B}}$ is neither exactly parallel nor perpendicular to $\bm{F}$, the stable fixed point shifts off of the north pole ($\theta^* \neq 0$). 
Note that $\theta^* = \theta^* (\theta_B)$ and $\phi^* = \phi^* (\phi_B)$, with the exception of the case $\theta^* = 0$, in which case $\phi^*$ is not defined.
In this perturbative regime, we find that the numerical results are summarized by
\begin{align}
\label{eq:fp_dep_B}
\theta^* = \epsilon A \sin{(k \, \theta_B)} 
\quad
\text{and}
\quad
\phi^* = \phi_B - \pi
\end{align}
where $A/\pi \simeq 0.0073$ and $k \simeq 2$ for $0 < \theta_B < \pi/2$. Increasing $\epsilon$ moves the fixed point further from the north pole.
If $\pi/2 < \theta_B < \pi$, the $\theta^*$ dependence remains the same as shown in Fig.~\ref{fig:numerics}d, and $\phi^*$ shifts by $\pi$.

With the help of Eq.~\ref{eq:fp_dep_B}, it is possible to adiabatically change the axis of $\bm{B}$ over time to induce the orientation of the pushpin to follow some desired trajectory.
Figure~\ref{fig:control} provides the necessary protocol for $\theta_B(t)$ and $\phi_B(t)$ that drives the pushpin to rotate in such a way as to trace out a rose curve, which has the form $r = d \cos{(m \phi)}$ in polar coordinates (here $m = 4$).
The control loop here is open: the axis of $\bm{B}$ affects the orientation of the pushpin, but there is no feedback on $\bm{B}$ from the current orientation.

We now introduce a simplified description of the orientation dynamics.
In the isotropic case, the orientation vector field $\bm{N}$ in Eq.~\ref{eq:N}-\ref{eq:eom_spherical} is well-approximated by $\bm{N}_{\text{iso}} = (0, -\sin{\theta}, 0)$ in spherical coordinates (Fig.~\ref{fig:setup}c). 
From this, we can construct a toy model of $
\bm{N}$ in the case $\epsilon \neq 0$. 
To obtain the flow to a fixed point ($\theta^*, \phi^*$) which is off of the north pole, we can simply rotate the isotropic vector field to find
\begin{equation}
\!\!\!\!\!\!\bm{N}_{\text{an}} =
\begin{bmatrix}
0\\
\cos{\theta} 
\sin{\theta^*}\cos{(\phi^* - \phi)}  - \cos{\theta^*}\sin{\theta}\\
\sin{\theta^*}\sin{(\phi^* - \phi)}
\end{bmatrix}
\!\!\!\!
\end{equation}
in spherical coordinates,
as shown in Fig.~\ref{fig:setup}d.
Note that in a few cases, the orientation dynamics can be reduced to a flow on a circle, for example if we constrain $\theta^* = \theta = \pi/2$, which corresponds to flow on the equator.

We achieve open loop control in the same way as in the full system: the fixed point ($\theta^*, \phi^*$) (the control variable) is simply set to the desired target ($\theta_{\text{set}}, \phi_{\text{set}}$) (Fig.~\ref{fig:control_loops}a).
In the presence of slowly varying noise, the orientation ($\theta(t), \phi(t)$) evolves through Eq.~\ref{eq:eom_spherical} to be near the target, but does not follow it exactly (Fig.~\ref{fig:control_loops}b). 
To improve control over the pushpin orientation, we close the feedback loop with a proportional-integral-derivative (PID) controller \cite{bechhoefer2021control} (Fig.~\ref{fig:control_loops}c-e) by setting
\begin{align}
    \begin{bmatrix}
    \theta^*\\
    \phi^*
    \end{bmatrix} (t)
    =
    K_p e(t) + K_i \int_0^t e(\tau) d\tau + K_d \frac{de(t)}{dt}
    \label{eq:PID}
\end{align}
where the error $e(t) = (\theta_{\text{set}}(t) - \theta^*(t), \phi_{\text{set}}(t)-\phi^*(t))$, and $K_p, K_i$, and $K_d$ are parameters of the PID controller.
Practically, we differentiate Eq.~\ref{eq:PID} with respect to time to obtain a set of ordinary differential equations, which we numerically solve in conjuction with Eq.~\ref{eq:eom_spherical} with the forward Euler method.
We find that the closed loop successfully controls the orientation (Fig.~\ref{fig:control_loops}d, compare with Fig.~\ref{fig:control_loops}b) by changing the fixed point (Fig.~\ref{fig:control_loops}e) in response to the noise (Fig.~\ref{fig:control_loops}f).

\textit{Discussion.}---Our work suggests a novel method of indirect control of objects through the modulation of the properties of the medium in which they are immersed. By changing the viscosity of an anisotropic fluid, we manipulate the orientation of a small particle that sediments under the force of gravity. Such control could be experimentally realized in anisotropic fluids by varying the alignment axis of the fluid molecules.

\medskip
The computer code used in this study is available on Zenodo at \cite{khain_2024_13308414} under the 2-clause BSD license.

\medskip

\begin{acknowledgments}
\textbf{Acknowledgements.} We thank Tom Witten, Colin Scheibner, Bryan VanSaders, and Yael Avni for discussions.
T.K. acknowledges partial support from the National Science Foundation Graduate Research Fellowship under grant no. 1746045. M.F. acknowledges partial support from the National Science Foundation under grant no. DMR-2118415, a Kadanoff-Rice fellowship funded by the National Science Foundation under award no. DMR-2011854 and the Simons Foundation. V.V. acknowledges partial support from the Army Research Office under grant nos. W911NF-22-2-0109 and W911NF-23-1-0212 and the Theory in Biology program of the Chan Zuckerberg Initiative. M.F. and V.V. acknowledge partial support from the France Chicago centre through a FACCTS grant. This research was partly supported by the National Science Foundation through the Physics Frontier Center for Living Systems (grant no. 2317138) and the University of Chicago Materials Research Science and Engineering Center (award no. DMR-2011854).
\end{acknowledgments}

\appendix
\clearpage

\renewcommand{\theequation}{S\arabic{equation}}

\section{Appendix A: Anisotropic viscosities}
\label{app:Stokes_general}

A passive anisotropic fluid with cylindrical symmetry can contain three independent shear viscosity coefficients.
These can be obtained by explicitly writing down the transformation law for the viscosity tensor (see for instance Refs.~\cite{khain2022stokes} and references therein). At steady state, the Stokes equation yields
\begin{align}
        0 = -\nabla P 
        &+\mu_1
    \begin{bmatrix}
    (\partial_x^2 + \partial_y^2) v_x\\
    (\partial_x^2 + \partial_y^2) v_y\\
    0
    \end{bmatrix} \nonumber \\
    &+
    \mu_2
    \begin{bmatrix}
    \partial_z^2 v_x + \partial_x \partial_z v_z\\
    \partial_z^2 v_y + \partial_y \partial_z v_z\\
    (\partial_x^2 + \partial_y^2)v_z - \partial_z^2 v_z
    \end{bmatrix} \nonumber \\
    &+
    \mu_3
    \begin{bmatrix}
    -\partial_x \partial_z v_z\\
    -\partial_y \partial_z v_z\\
    2 \partial_z^2 v_z
    \end{bmatrix}.
    \label{eq:Stokes_full}
\end{align}
If $\mu_1 = \mu_2 = \mu_3 \equiv \mu$, the viscous contribution reduces to the familiar $\mu \Delta \bm{v}$. In the main text, we consider the case $\mu_1 = \mu, \mu_2 = \mu(1 + \epsilon)$, and $\mu_3 = \mu(1 + \frac{4}{3}\epsilon)$, where $\epsilon$ is small. In this case, the above equation reduces to Eqs.~\ref{eq:Stokes_anisotropic}-\ref{eq:Stokes_Bz}.
In a generic fluid, additional viscosity coefficients can be present, see Refs.~\cite{khain2022stokes} and references therein for more details.
The viscosities in Eq.~\ref{eq:Stokes_full} can be expressed in terms of the Leslie viscosity coefficients $\alpha_i$ (Eq.~6.50 of \cite{kleman2003soft}) as 
\begin{align}
\mu_1 &= \frac{\alpha_4}{2}\\
\mu_2 &= \frac{\alpha_4}{2} + \frac{\alpha_5 + \alpha_6}{4} \\
\mu_3 &= \frac{\alpha_4}{2} + \frac{\alpha_1 + \alpha_5 + \alpha_6}{3}
\end{align}
in which we have considered a uniform nematic director $\bm{n} = \hat{\bm{z}}$.

\section{Appendix B: Green function (Stokeslet)}
\label{app:Greens_function}

We compute the Green function (Stokeslet) corresponding to Eqs.~\ref{eq:Stokes_anisotropic}-\ref{eq:Stokes_Bz} in the perturbative regime, for small anisotropy in the $z$ direction ($\hat{\bm{B}} = \hat{\bm{z}}$).
The general case (for arbitrarily large $\epsilon$) can be computed numerically with fast Fourier transforms. The Stokeslet is the solution to the Stokes equation with an applied point force~$\bm{F}$:
\begin{align}
\bm{F}\delta^3(\bm{r}) =  -\nabla P + \mu\Delta \bm{v} + \epsilon \mu
\begin{bmatrix}
    \partial_z^2 v_x - \partial_x \partial_z v_z/3\\
    \partial_z^2 v_y - \partial_y \partial_z v_z/3\\
    \Delta v_z + 2\partial_z^2 v_z/3
\end{bmatrix}
\label{eq:Stokeslet_eq_special}
\end{align}

where we take $\epsilon \ll 1$. To solve for $\bm{v}$, we write Eq.~\ref{eq:Stokeslet_eq_special} in Fourier space, solve for the pressure and velocity fields, and integrate using contour integration to find the real-space solutions, in the same way as in \cite{khain2022stokes}.

The Stokeslet velocity field is expressed through the Green function, $\mathbb{G}$, as $\bm{v}(\bm{r}) = \mathbb{G}(\bm{r}) \bm{F}$.
Expanding the Green function to linear order in $\epsilon$, $\mathbb{G}(\bm{r}) = \mathbb{G}_0(\bm{r}) + \epsilon \mathbb{G}_1(\bm{r})$, we recover the familiar solution for a normal fluid,
\begin{align}
\mathbb{G}_{0,ij}(\bm{r}) &= \frac{1}{8\pi \mu r^3}\left(r^2\delta_{ij} + r_i r_j\right)
\end{align}
and derive the first order correction due to anisotropy, $\mathbb{G}_1(\bm{r})$:
\begin{align}
\mathbb{G}_1(\bm{r}) = \frac{1}{8\pi \mu r^3}
    \begin{bmatrix}
        -(x^2 + y^2) & 0 & -xz\\
        0 & -(x^2 + y^2) & -yz\\
        -xz & -yz & -(x^2 + y^2 + 2z^2)
    \end{bmatrix}.
    \label{eq:G_Bz}
\end{align}
The associated pressure field is $P(\bm{r}) = P_0(\bm{r}) + \epsilon P_1(\bm{r})$, with
\begin{align}
P_0(\bm{r}) &= \frac{1}{4\pi r^3} \bm{F}\cdot \bm{r}\\
P_1(\bm{r}) &= -\frac{x^2 + y^2 - 2z^2}{12\pi r^5}(\bm{F}\cdot \bm{r} - 2 \bm{F}\cdot \bm{z}).
\end{align}

The Green function in Eq.~\ref{eq:G_Bz} holds for an anisotropy axis $\hat{\bm{B}} = \hat{\bm{z}}$. 
Under a rotation to an arbitrary anisotropy axis given by $\hat{\bm{B}} = (\cos{\phi_B}\sin{\theta_B}, \sin{\phi_B}\sin{\theta_B}, \cos{\theta_B})$, the Green function in Eq.~\ref{eq:G_Bz} transforms as
\begin{equation}
\mathbb{G}_1(\bm{r}) \to R\, \mathbb{G}_1(R^{-1}\bm{r})R^{-1},
\label{eq:G_general}
\end{equation}
where $R$ is the rotation matrix
\begin{equation}
R = 
    \begin{bmatrix}
    \cos{(\theta_B)} \cos{(\phi_B)} & - \sin{(\phi_B)} & \cos{(\phi_B)} \sin{(\theta_B)} \\
    \cos{(\theta_B)} \sin{(\phi_B)} & \cos{(\phi_B)} & \sin{(\phi_B)} \sin{(\theta_B)} \\
    -\sin{(\theta_B)} & 0 & \cos{(\theta_B)}
    \end{bmatrix}.
    \label{eq:rot_matrix}
\end{equation}

\section{Appendix C: Flow past a sphere}
\label{app:flow_sphere}
We solve the anisotropic Stokes equation (Eqs.~\ref{eq:Stokes_anisotropic}-\ref{eq:Stokes_Bz}) for the flow past a sphere to linear order in $\epsilon$ by writing $\bm{v}(\bm{r}) = \bm{v_0}(\bm{r}) + \epsilon \bm{v_1}(\bm{r}), P(\bm{r}) = P_0(\bm{r}) + \epsilon P_1(\bm{r})$, as in \cite{khain2022stokes}. 
We take the velocity of the fluid at infinity to be $\bm{U}$, and the boundary condition on the sphere surface to be no-slip, $\bm{v}(r = a) = 0$, where $a$ is the sphere radius. 

Since the viscosity is anisotropic, we have two cases to consider: one in which $\bm{U}$ is parallel to the anisotropy axis $\bm{B}$, and one in which $\bm{U}$ and $\bm{B}$ are perpendicular.

Let us take $\hat{\bm{B}} = \hat{\bm{z}}$. We first consider the parallel case, $\bm{U} = U \hat{\bm{z}}$. In this situation, the velocity field around the sphere is not modified at first order, $\bm{v_1}(\bm{r}) = 0$, but the pressure is:
\begin{align}
P_{1}(\bm{r}) &= -\frac{\mu a U}{2r^7} z(4x^4 + 4y^4 + 5y^2 z^2 + z^4 \nonumber\\
&+ 8x^2y^2 + 5x^2 z^2 + a^2(-3x^2 - 3y^2 + 2z^2)).
\end{align}

We repeat in the perpendicular case, for $\bm{U} = U \hat{\bm{x}}$. Here, both the velocity and pressure fields are modified,
\begin{align}
v_{1,x}(\bm{r}) &= \frac{3aU}{8r^5}(y-z)(y+z)(r^2 - a^2)\\
v_{1,y}(\bm{r}) &= -\frac{3aU}{8r^5}xy(r^2 - a^2)\\
v_{1,z}(\bm{r}) &= \frac{3aU}{8r^5}xz(r^2 - a^2)\\
P_1(\bm{r}) &= \frac{\mu U a}{4r^7} x (-5(x^2 + y^2)^2 - 4(x^2 + y^2)z^2 \nonumber\\
&+ z^4 + 2a^2 (x^2 + y^2 -4z^2)) 
\end{align}
The velocity field can be more compactly written in terms of the Green function and its Laplacian. 
To linear order, the velocity is
\begin{align}
\bm{v}(\bm{r}) = &-6\pi \mu U a \left(1 + \frac{\epsilon}{2}\right)\mathbb{G}(\bm{r}) \cdot \hat{\bm{x}} \nonumber \\ 
&-\pi \mu U a^3 \left(1 + \frac{\epsilon}{2}\right) \Delta \mathbb{G}(\bm{r})\cdot \hat{\bm{x}}.
\end{align}
The first order term can be written explicitly as
\begin{align}
\bm{v_1}(\bm{r}) = &-6\pi \mu U a\left(\frac{\mathbb{G}_0(\bm{r})}{2} \cdot \hat{\bm{x}} + \mathbb{G}_1(\bm{r}) \cdot \hat{\bm{x}}\right) \nonumber \\
&-\pi \mu U a^3\Delta \left(\frac{\mathbb{G}_0(\bm{r})}{2} \cdot \hat{\bm{x}} + \mathbb{G}_1(\bm{r}) \cdot \hat{\bm{x}}\right).
\end{align}

\section{Appendix D: Forces on a sphere}
\label{app:drag_sphere}
To solve for the forces on the sphere due to the fluid flow, we compute the stress from the above velocity fields and integrate it over the surface of the sphere,
\begin{equation}
F_i = \oint \sigma_{ij} n_j dS,
\end{equation}
where $\hat{\bm{n}} = \hat{\bm{r}}$ is the unit vector normal to the sphere surface.

Here, in addition to the familiar pressure and shear viscosity contributions, the stress contains a third term due to the anisotropic viscosity,
\begin{equation}
\sigma_{ij} = -P\delta_{ij} + \mu(\partial_i v_j + \partial_j v_i) + \epsilon \sigma_{\text{an}},
\end{equation}
where
\begin{widetext}
\begin{align}
\sigma_{\text{an}} = \mu
    \begin{bmatrix}
    \frac{4}{9} (\partial_x v_x + \partial_y v_y - 2\partial_z v_z) & 0 & \partial_z v_x + \partial_x v_z\\
    0 & \frac{4}{9}(\partial_x v_x + \partial_y v_y - 2\partial_z v_z) & \partial_z v_y + \partial_y v_z\\
    \partial_z v_x + \partial_x v_z & \partial_z v_y + \partial_y v_z & -\frac{8}{9} (\partial_x v_x + \partial_y v_y - 2\partial_z v_z)
    \end{bmatrix}
\end{align}
\end{widetext}
for an anisotropy axis $\hat{\bm{B}} = \hat{\bm{z}}$.

Computing the forces yields the following subset of the propulsion matrix:
\begin{align}
    \begin{bmatrix}
        F_x\\
        F_y\\
        F_z
    \end{bmatrix}
    = 6\pi\mu a
    \begin{bmatrix}
        1+\frac{\epsilon}{2} & 0 & 0\\
        0 & 1+\frac{\epsilon}{2} & 0 \\
        0 & 0 & 1 + \epsilon
    \end{bmatrix}
    \begin{bmatrix}
        V_x \\
        V_y \\
        V_z
    \end{bmatrix}.
\end{align}
Due to the anisotropy of the viscosity, the Stokes drag law is modified. The drag coefficients in the $x$ and $y$ directions are increased by a factor of $(1 + \epsilon/2)$ and in the $z$ direction (along the anisotropy axis) by $(1 + \epsilon)$. The fluid does not exert torques on the sphere.

The $A$ block of the mobility matrix $\mathbb{M}$ (see Eq.~\ref{eq:mobility_blocks}) is simply the inverse of the matrix above:
\begin{align}
    \begin{bmatrix}
        V_x\\
        V_y\\
        V_z
    \end{bmatrix}
    = 
    \underbrace{
    \frac{1}{6\pi\mu a}
    \begin{bmatrix}
        1-\frac{\epsilon}{2} & 0 & 0\\
        0 & 1-\frac{\epsilon}{2} & 0 \\
        0 & 0 & 1 - \epsilon
    \end{bmatrix}
    }_A
    \begin{bmatrix}
        F_x \\
        F_y \\
        F_z
    \end{bmatrix}.
    \label{eq:sphere_mobility}
\end{align}

Eq.~\ref{eq:sphere_mobility} holds for an anisotropy axis $\hat{\bm{B}} = \hat{\bm{z}}$.
To obtain Eq.~\ref{eq:sphere_mobility} for an arbitrary anisotropy axis $\hat{\bm{B}} = (\cos{\phi_B}\sin{\theta_B}, \sin{\phi_B}\sin{\theta_B}, \cos{\theta_B})$, we transform $A$ as follows:
\begin{equation}
A \to R A R^{-1},
\label{eq:M_general}
\end{equation}
where $R$ is the rotation matrix in Eq.~\ref{eq:rot_matrix}.
Note that we can transform $\mathbb{M}$ in this way only for the sphere due to its rotational invariance.
For the pushpin, which has its own anisotropy axis $\bm{n}$, $\mathbb{M}$ only transforms as in Eq.~\ref{eq:M_general} if $\bm{B}$ and $\bm{n}$ rotate together.
In the general case, the mobility matrix must be recomputed for different anisotropy axes, as described below.

\section{Appendix E: Stokeslet approximation of the mobility matrix}
\label{app:Stokeslet_approx}

To isolate the coefficients that relate different degrees of freedom, the mobility matrix can be conveniently arranged in four blocks
\begin{align}
    \mathbb{M} =
    \begin{bmatrix}
        \begin{array}{@{}c|c@{}}
        A & B \\
        \hline
        T & S 
        \end{array}
    \end{bmatrix}.
    \label{eq:mobility_blocks}
\end{align}

For shapes that are less symmetric than the sphere, it is difficult to obtain an analytical form of the mobility matrix. 
To derive the mobility matrix of the pushpin-shaped object, we construct the pushpin out of small spheres of radius $a$ (denoted by markers in Fig.~\ref{fig:numerics}a) with the method reviewed in \cite{witten2020review},  see \cite{khain2024trading, khain_2024_13308414} for our implementation.

With this method, we apply a force to the pushpin at some reference point, which is then distributed amongst the small spheres.
Reference \cite{witten2020review} provides an algorithm to determine how to distribute these forces that depends on two main ingredients.

The first ingredient is the velocity field generated by a small sphere moving in a fluid due to an applied force.
The distance between the spheres that compose the pushpin is taken to be much larger than the radius of the spheres, which allows us to treat the spheres as Stokeslets, and approximate the velocity field by the Green function in Eq.~\ref{eq:G_Bz}.
The second ingredient is the force on a sphere moving with some velocity in a fluid (i.e. the $A$ block of the mobility matrix), which we derived in Eq.~\ref{eq:sphere_mobility}.

Combining these two ingredients, we impose the constraint of a rigid body (we insist that the small spheres cannot move relative to one another) which yields the mobility matrix of the pushpin.
Moreover, with the help of Eqs.~\ref{eq:G_general} and~\ref{eq:M_general}, we can compute the mobility matrix of the pushpin for any orientation of the anisotropy axis $\hat{\bm{B}}$.

For the computations in this work, the pushpin is composed of fourteen spheres of radius $a = 0.01$. The four which lie along the axis are spaced with unit distance, and the ten that are along the base lie on the vertices of a regular decagon. The line segments that connect the markers in Fig.~\ref{fig:numerics}a are not real and are meant to guide the eye.

Below, we provide the numerical values of $T$, the bottom left block of the mobility matrix shown pictorially in Eq.~\ref{eq:M_orient}. For these matrices, the pushpin orientation is $\phi = 0, \theta = 0.3\pi$. For the anisotropic case, we take $\epsilon = 0.1$, $\hat{\bm{B}} = \frac{1}{\sqrt{2}}(0, 1, 1)$.
\begin{align}
    \!\!\!T_{\text{iso}}&\simeq
    \begin{bmatrix}
    0 & 0.00367034 & 0\\
    -0.00367034 & 0 & 0.00505179\\
    0 & -0.00505179 & 0
    \end{bmatrix}\\
    \!\!\!T_{\text{an}}&\simeq
    \begin{bmatrix}
    0.00007423 & 0.00345387 & -0.00010948\\
    -0.00350251 & -0.00020758 & 0.00481074\\
    -0.00008957 & -0.00473651 & 0.00013335
    \end{bmatrix}
\end{align}

\providecommand{\noopsort}[1]{}\providecommand{\singleletter}[1]{#1}

\end{document}